\documentclass{article}
\usepackage{jheppub}
\usepackage{graphicx} %
\usepackage{amsmath,amssymb,commath,braket}
\usepackage[most]{tcolorbox}
\usepackage{mathtools}
\usepackage{cleveref}
\usepackage{bbm}
\usepackage{caption}
\usepackage{subcaption}
\usepackage{soul}
\usepackage{cancel}
\usepackage{orcidlink}
\usepackage{placeins}

\DeclareMathOperator{\Tr}{Tr}

\newcommand{\hm}[1]{\textcolor{blue}{(HM: #1)}}

\title{Phase Transitions at More Unusual Values of Theta}

\author[a,b]{Dhruv Aldas \orcidlink{0009-0007-0103-1174}}\emailAdd{dhruv\_aldas@berkeley.edu}
\author[a,b,c,1]{Hitoshi Murayama\,\orcidlink{0000-0001-5769-9471},\note{Hamamatsu Professor}}\emailAdd{hitoshi@berkeley.edu}
\author[a,b]{Bea Noether\, \orcidlink{0000-0002-2947-3210},}\emailAdd{bea\_noether@berkeley.edu}
\author[a,b]{Digvijay Roy Varier\,\orcidlink{0000-0002-4695-7428}}\emailAdd{digvijayroyvarier@berkeley.edu}

\affiliation[a]{Leinweber Institute for Theoretical Physics, University of California, Berkeley, CA 94720, USA}
\affiliation[c]{Kavli Institute for the Physics and Mathematics of the Universe (WPI), University of Tokyo, Kashiwa 277-8583, Japan}
\affiliation[b]{Ernest Orlando Lawrence Berkeley National Laboratory, Berkeley, CA 94720, USA}

\abstract{
We calculate the $\theta$ dependence of the vacuum energy in an infinite family of cousins of QCD, where the vacuum structure can be analyzed exactly. These theories are $\mathcal{N}=1$ EFTs that descend from $\mathcal{N}=2$ $SU(N_c)$ gauge theory with $N_f$ flavors of fundamental hypermultiplets, that are then deformed by anomaly mediated supersymmetry breaking (AMSB). We reproduce results found in previous work that used the full $\mathcal{N}=2$ machinery but via a much simpler $\mathcal{N}=1$ method. Our method easily extends to infinitely many more theories for which the full $\mathcal{N}=2$ approach becomes intractable. We identify values of $\theta$ for which the theory undergoes a phase transition, and find that they can occur at both fractional and irrational values of $\theta/\pi$.
}

\begin{document}
\maketitle

\section{Introduction}\label{sec:intro}

The dependence of gauge theories on the topological angle $\theta$ is one of the sharpest windows into their nonperturbative vacuum structure. It underlies the strong CP problem, controls the multiplicity and energetics of vacua, and is tightly constrained by anomalies and the periodicity $\theta\sim\theta+2\pi$. In pure Yang--Mills and QCD, general arguments (Dashen's phenomenon \cite{dashen_features_1971,hooft_topology_1981,cardy_phase_1982}, large-$N_c$ reasoning, and more recently the mixed 't~Hooft anomaly between CP and the center at $\theta=\pi$ \cite{gaiotto_theta_2017}) indicate that the vacuum energy is a nontrivial, multi-branched function of $\theta$~\cite{witten_large_n_1980,witten_theta_1998}, with candidate phase transitions at the CP-symmetric points $\theta=0,\pi$ \cite{konishi_confinement_1997,evans_phase_1997}. Away from these special points, however, direct information is scarce: the theory is strongly coupled and the vacuum energy is not calculable from first principles. It is therefore valuable to have exactly solvable ``cousins of QCD'', theories with the same gauge group and matter content but with additional structure (here, softly broken supersymmetry) that renders the vacuum energy computable, in which the full $\theta$-dependence, including transitions at fractional and irrational values of $\theta/\pi$, can be mapped out explicitly.

Such a laboratory was recently exploited in Ref.~\cite{csaki_phase_2025}. There, an $\mathcal N=2$ analogue of QCD, an $SU(N_c)$ theory with $N_f$ fundamental hypermultiplets, was broken to $\mathcal N=1$ by an adjoint mass and then further perturbed by anomaly-mediated supersymmetry breaking (AMSB). Using the full $\mathcal N=2$ special geometry, the authors computed the $\theta$-dependence of the vacuum energy and found phase transitions at fractional values of $\theta/\pi$ (see also \cite{konishi_confinement_1997,evans_phase_1997} for earlier work). Their analysis, however, was carried out only for $N_c=2$ and $N_f<4$. One would like to extend it to larger numbers of colors and flavors, but the Seiberg--Witten special geometry~\cite{seiberg_witten_1994a, seiberg_witten_1994b} underlying the $\mathcal N=2$ computation rapidly becomes intractable as $N_c$ grows~\cite{dhoker_multimonopole_2021, dhoker_strong_coupling_2022}.

In this paper we show that the same vacuum dependence and critical angles can be obtained from a much simpler starting point. Rather than solving the $\mathcal N=2$ theory exactly, we begin directly from the $\mathcal N=1$ low-energy effective description: $\mathcal N=1$ SQCD$(N_c,N_f)$ deformed by the quartic meson superpotential $\Delta W=-\frac{1}{\mu}(\tilde{Q}T^a Q)(\tilde{Q} T^a Q)$, obtained by integrating out the massive adjoint, and then breaking to $\mathcal{N}=0$ by AMSB. The AMSB spurion lifts the degeneracy among the discrete supersymmetric vacua, and since supersymmetric extrema have vanishing $F$-term energy, the leading $\theta$-dependence is captured entirely by the compensator-induced terms. This reduces the determination of the vacuum energy to a branch-counting problem governed by the holomorphic phases of the strong-coupling scales, and it extends immediately to generic $(N_c,N_f)$. 

The effective $\mathcal N=1$ description is manifestly valid in the weakly-coupled region of the moduli space, whereas Ref.~\cite{csaki_phase_2025} perturbs the strongly-coupled dyon point. Nevertheless, holomorphy prevents phase transitions as superpotential parameters are varied in supersymmetric theories \cite{Seiberg:1994bp}, since a phase boundary cannot be expressed as a holomorphic condition. The $\mathcal N=2$ and $\mathcal N=1$ descriptions correspond to different orderings of the two holomorphic parameters, the $\mathcal N=2$ dynamical scale $\Lambda_{\mathcal{N}=2}$ and the adjoint mass $\mu$. Thus, in the absence of AMSB, smooth interpolation between the two regimes is expected. Our results suggest that this continuity persists for sufficiently small AMSB, provided $m\ll\mu,\Lambda_{\mathcal{N}=2}$. The $\mathcal N=1$ effective description is parametrically controlled for $m\ll\Lambda_{\mathcal{N}=2}\ll\mu$.

Moreover, as explained in Sec.~\ref{sec:AMSB}, for these theories the determination of the vacuum energy to leading order in $m$ is independent of the K\"ahler potential. As such our results follow purely from the superpotential and are insensitive to any unknown features of the K\"ahler potential that could pose obstructions in other contexts.

Our main results are explicit expressions for the AMSB vacuum energy and the resulting critical angles $\theta_{\rm crit}$ across the four dynamical regimes of SQCD with $N_f/N_c<3/2$. In each case, the vacuum energy at $\mathcal{O}(m)$ reduces to either zero or a cosine in $\theta$ with variable amplitude and phase shift, and $\theta_{\rm crit}$ is fixed by discrete, parity-dependent data ($N_c$, $N_f$, and combinations such as $2N_c-N_f$). For $N_c=2$, these vacua agree with those found using the full $\mathcal N=2$ machinery of Ref.~\cite{csaki_phase_2025}. Consistent with the holomorphic continuity described above, this agreement motivates the central conjecture of this paper. The $\mathcal N=2$ and $\mathcal N=1$ computations correspond to two orderings of the same operations: (i) flow to low energy within $\mathcal N=2$, then break to $\mathcal N=1$ and add AMSB, or (ii) break to $\mathcal N=1$ and add AMSB first, then flow to low energy. We conjecture that the two orderings yield identical IR theories.

The restriction $N_f/N_c<3/2$ deserves comment. For $N_f/N_c>3/2$ the theory enters the conformal window, where explicit results are available only in the large-$N_c$ limit at fixed $N_f/N_c$~\cite{kondo_broken_2025,csaki_guide_2023,seiberg_duality_1995}. Since our critical angles depend sensitively on the parity of $N_c$, $N_f$, or $2N_c-N_f$, and such a large-$N_c$ limit washes out any definite parity, we do not expect sharp predictions for $\theta_{\rm crit}$ in the conformal window and accordingly restrict to $N_f/N_c<3/2$.

The remainder of the paper is organized as follows. In Sec.~\ref{sec:N=2Breaking} we describe the $\mathcal N=2\to\mathcal N=1$ deformation and re-express the quartic deformation in terms of gauge invariants. In Sec.~\ref{sec:AMSB} we review AMSB and the spurion prescription we use throughout. We then treat the four dynamical regimes in turn, each governed by distinct low-energy physics: the Affleck--Dine--Seiberg superpotential for $N_f<N_c$ (Sec.~\ref{sec:ADS}); the quantum-modified moduli space for $N_f=N_c$ (Sec.~\ref{sec:QM}); s-confinement for $N_f=N_c+1$ (Sec.~\ref{sec:sconf}); and the free magnetic (Seiberg dual) description for $N_c+1<N_f<\tfrac32 N_c$ (Sec.~\ref{sec:FreeMag}). We conclude in Sec.~\ref{sec:conclusion}.

\section{Breaking $\mathcal{N}=2$ to $\mathcal{N}=1$}\label{sec:N=2Breaking}
We consider $\mathcal{N}=1$ SQCD($N_c$,$N_f$) deformed by the superpotential $\Delta W = -\frac{1}{\mu}(\tilde{Q}T^a Q)(\tilde{Q} T^a Q)$ and AMSB. This is the effective $\mathcal{N}=1$ theory obtained by taking $\mathcal{N}=2$ $SU(N_c)$ gauge theory with $N_f$ fundamental hypermultiplets and deforming it by the superpotential $W=\mu \Tr\Phi^2 \equiv \frac{\mu}{2} \Phi^a \Phi^a$ (in addition to the $\mathcal{N}=2$ superpotential), where $\Phi$ is the adjoint chiral superfield~\cite{seiberg_witten_1994a, seiberg_witten_1994b, argyres_plesser_seiberg_1996}. Upon integrating out $\Phi$ we obtain $\mathcal{N}=1$ SQCD($N_c,N_f$) deformed by $\Delta W$, as stated above.
This process agrees with the deformation in Ref.~\cite{csaki_phase_2025} in the weakly coupled regime. That reference instead perturbs the strongly coupled dyon description. As discussed in the Introduction, holomorphy motivates a smooth interpolation between these regimes in the supersymmetric limit, while our calculation indicates that this remains true after introducing sufficiently small AMSB.

Note that $\Delta W$ can be re-expressed in terms of gauge invariants. Consider the composite operator
\begin{equation}\label{eq:O_def}
    O = \left[\tilde{Q}_i^\alpha \, (T^a)_\alpha^{\ \beta} \, Q_\beta^i \right]\; \left[\tilde{Q}_j^\gamma \, (T^a)_\gamma^{\ \delta} \, Q_\delta^j \right]\,,
\end{equation}
where $T^a$, $a=1,\dots,N_c^2-1$, are the generators of $SU(N_c)$ in the fundamental representation. We wish to express $O$ in terms of the meson matrix $M_i{}^j\equiv \tilde{Q}_i^\alpha Q^j_\alpha$.

\medskip

The key ingredient is the Fierz (completeness) identity for the generators of $SU(N_c)$:
\begin{equation}\label{eq:fierz}
    (T^a)_\alpha^{\ \beta} \, (T^a)_\gamma^{\ \delta} = \frac{1}{2}\left( \delta_\alpha^{\ \delta} \, \delta_\gamma^{\ \beta} - \frac{1}{N_c} \, \delta_\alpha^{\ \beta} \, \delta_\gamma^{\ \delta} \right) .
\end{equation}
Substituting~\eqref{eq:fierz} into~\eqref{eq:O_def}, we obtain
\begin{equation}
    O = \frac{1}{2} \, \tilde{Q}_i^\alpha \, Q_\beta^i \; \tilde{Q}_j^\gamma \, Q_\delta^j \left( \delta_\alpha^{\ \delta} \, \delta_\gamma^{\ \beta} - \frac{1}{N_c} \, \delta_\alpha^{\ \beta} \, \delta_\gamma^{\ \delta} \right) .
\end{equation}
We now contract the color indices in each term separately. For the first term, contracting with $\delta_\alpha^{\ \delta} \, \delta_\gamma^{\ \beta}$ gives
\begin{equation}
    \frac{1}{2} \, \tilde{Q}_i^\alpha \, Q_\gamma^i \; \tilde{Q}_j^\gamma \, Q_\alpha^j = \frac{1}{2} \, M_i^{\ j} \, M_j^{\ i} = \frac{1}{2} \Tr(M^2) \,.
\end{equation}
For the second, contracting with $\delta_\alpha^{\ \beta} \, \delta_\gamma^{\ \delta}$ gives
\begin{equation}
    -\frac{1}{2N_c} \, \tilde{Q}_i^\alpha \, Q_\alpha^i \; \tilde{Q}_j^\gamma \, Q_\gamma^j = -\frac{1}{2N_c} \, M_i^{\ i} \, M_j^{\ j} = -\frac{1}{2N_c}(\Tr M)^2 .
\end{equation}
Combining both contributions, we arrive at the result:
\begin{equation}
    O = \frac{1}{2}\Tr(M^2) - \frac{1}{2N_c}(\Tr M)^2
\end{equation}
This is the form of the deformation we will use moving forward. Note that this seemingly weakly coupled analysis of composite operators is allowed because there are no short-distance singularities among chiral superfields\footnote{This is true at all orders in perturbation theory. However non-perturbative corrections are possible in gauge-invariant invariant operators that can cause some ambiguities. See discussions on $N_f=N_c$ in \cref{sec:QM}.} \cite{Lerche:1989uy,Cachazo:2002ry}.

It is important to note the difference between the $\mathcal{N}=1$ and $\mathcal{N}=2$ holomorphic scales. Namely:
\begin{align}
    \Lambda_{\mathcal{N}=1}^{3N_c-N_f} =&~ \mu^{N_c}\Lambda_{\mathcal{N}=2}^{2N_c-N_f}. \label{eq:scalematching}
\end{align}
In Ref.~\cite{csaki_phase_2025} several different shorthands for combinations of $\Lambda_{\mathcal{N}=2}$ and constants are introduced, and notably the scale that appears in the Seiberg--Witten curve equations is normalized differently. The convention can be matched by considering the result for pure SYM. For AMSB-deformed $\mathcal{N}=1$ SYM the superpotential from gaugino condensation is $W=N_c\Lambda_{\mathcal{N}=1}^3$ and as such the vacuum energy to leading order is $V = -6m N_c\abs{\Lambda_{\mathcal{N}=1}}^3 \cos\left(\frac{\theta+2\pi k}{N_c}\right)$ for $k=0,...,N_c-1$. In Ref.~\cite{csaki_phase_2025} they write for $(N_c,N_f)=(2,0)$ that $V = -6m\mu\abs{\Lambda_0}^2\cos\left(\frac{\theta}{2}+(k-1)\pi\right)$ for $k=1,2$. This is equivalent if $\abs{\Lambda_0}^2 \mu = 2\abs{\Lambda_{\mathcal{N}=1}}^3$. It follows that their $\Lambda_0 = \Lambda$ is related to $\Lambda_{\mathcal{N}=2}$ (in the $(N_c,N_f)=(2,0)$ case) by $\abs{\Lambda_0}^2= 2\abs{\Lambda_{\mathcal{N}=2}}^2$. Similar considerations lead to the following dictionary:
$$\begin{array}{c|c|c}
N_f
& \text{Scale defined in \cite{csaki_phase_2025}}
& \text{Relation to }\Lambda_{\mathcal N=2}
\\[2mm]\hline
0
& \Lambda_0=\Lambda
& \Lambda_0=2^{1/2}\Lambda_{\mathcal N=2}
\\[2mm]
1
& \Lambda_1=\sqrt{3}\,2^{-4/3}\Lambda
& \Lambda_1=\sqrt{3}\,2^{-2/3}\Lambda_{\mathcal N=2}
\\[2mm]
2
& \Lambda_2=2^{-3/2}\Lambda
& \Lambda_2=2^{-1/2}\Lambda_{\mathcal N=2}
\\[2mm]
3
& \Lambda_3=2^{-4}\Lambda
& \Lambda_3=\frac14\Lambda_{\mathcal N=2}
\end{array}$$
We will see that when properly accounting for this normalization, we reproduce the vacuum energies of Ref.~\cite{csaki_phase_2025} exactly.

\subsection{Comparison with Ref.~\cite{csaki_phase_2025}}

We use the same adjoint-mass and theta-angle conventions as
Ref.~\cite{csaki_phase_2025}:
\begin{align}
W_{\rm UV}
=
\sqrt{2}\,\widetilde Q\Phi Q
+\mu\Tr\Phi^2,
\qquad
\mu>0.
\end{align}
Integrating out the adjoint gives
\begin{align}
\Delta W
=
-\frac{1}{\mu}
(\widetilde QT^aQ)(\widetilde QT^aQ)
=
-\frac{1}{2\mu}
\left[
\Tr(M^2)-\frac{1}{N_c}(\Tr M)^2
\right].
\end{align}
After applying the scale dictionary above, our $N_c=2$ branch
energies (special cases of the derivations in \cref{sec:ADS,sec:QM,sec:sconf}) are
\begin{align}
N_f=0:\qquad
V_k
&=
-6m\mu|\Lambda_0|^2
\cos\left[
\frac{\theta}{2}+(k-1)\pi
\right],
\\
N_f=1:\qquad
V_k
&=
+6m\mu|\Lambda_1|^2
\cos\left[
\frac{2(\theta+(k-2)\pi)}{3}
\right],
\\
N_f=2:\qquad
V_k
&=
-6m\mu|\Lambda_2|^2
\cos\left[
\theta+(k-1)\pi
\right],
\\
N_f=3:\qquad
V_1
&=
0+\mathcal O(m^2),
\\
V_2
&=
-6m\mu|\Lambda_3|^2
\cos(2\theta)
+\mathcal O(m^2).
\end{align}
These are precisely Eqs.~(4.7), (4.10), (4.13), and (4.18) of
Ref.~\cite{csaki_phase_2025}, up to harmless relabelings of the
discrete branches.

\section{Anomaly Mediated SUSY Breaking}\label{sec:AMSB}

In our broader AMSB program, the deformation by a small AMSB spurion $m$ is used as a controlled way to follow vacuum structure across strongly coupled regimes, and (when appropriate) to interpolate toward non-supersymmetric gauge dynamics~\cite{murayama_qcd_like_2021, csaki_guide_2023, csaki_chiral_2021, csaki_symmetric_tensor_2021, csaki_so_nc_2021, csaki_so_nc_case_2021, kondo_simplest_2022, leedom_flavor_2025,luty_soft_1999} (see~\cite{dine_yu_2022, dine_2022} for discussion on the feasibility of such extrapolation). In the present work, we use AMSB in a narrower sense: we start from the exact $\mathcal{N}=1$ low-energy effective theories obtained by deforming $\mathcal{N}=2$ SQCD, and we use AMSB to lift the discrete degeneracy among supersymmetric vacua so as to determine the $\theta$-dependence of the vacuum energy and locate values $\theta_{\rm crit}$ where the preferred branch changes. All of our statements are exactly derived features of the theories in question, and it is only an extrapolation toward the non-SUSY limit that would rely on conjecture.

AMSB~\cite{randall_out_1999, giudice_gaugino_1998, pomarol_sparticle_1999, arkani_hamed_giudice_1998, arkani_hamed_rattazzi_1999} is implemented by introducing a compensator superfield
\begin{equation}
\Phi=1+m\,\theta^2 \, ,
\end{equation}
which parametrizes the conformal-breaking spurion responsible for anomaly mediation. Expanding the resulting action in $m$ yields a universal tree-level contribution to the scalar potential~\cite{pomarol_sparticle_1999,csaki_guide_2023} that can be written in terms of the K\"ahler potential $K$ and superpotential $W$ of the rigid theory:
\begin{align}\label{eq:extendedAMSB}
V_{\rm tree} \;=\;&
\partial_i W\, g^{i j^*}\, \partial_{j^*} W^*
\;+\; m^* m \left( \partial_i K\, g^{i j^*} \partial_{j^*} K - K \right)\nonumber\\
&\quad+\; m \left(\partial_i W\, g^{i j^*}\, \partial_{j^*} K - 3 W \right)
+ c.c.\, ,
\end{align}
where $g_{ij^*}=\partial_i\partial_{j^*}K$ and $g^{ij^*}$ denotes its inverse. We take $m$ real in the following using the non-anomalous $U(1)_R$ rotation without a loss of generality that changes the phase of $\mu$. In addition, we can use the anomalous $U(1)_A$ to rotate $Q$ and $\tilde{Q}$ together to make $\mu$ real to absorb its phase into $\theta$. 

If $K$ is canonical in the variables used to describe the low-energy effective theory,~\eqref{eq:extendedAMSB} simplifies to the familiar form~\cite{randall_out_1999,giudice_gaugino_1998}
\begin{align}\label{eq:AMSBW}
V_{\rm tree} \;=\; m\left(\varphi_i \frac{\partial W}{\partial \varphi_i} - 3 W\right)+c.c.\, ,
\end{align}
where $\varphi_i$ are the scalar components of the chiral multiplets in $W$.

For our purposes, we want to find the leading-order vacuum energy in the small $m$ limit. The dependence on the K\"ahler potential drops out of this process at $\mathcal{O}(m)$. Namely, the scalar potential at this order is
\begin{align}
V_{\rm SUSY} + V_{\rm AMSB} =& 
g^{i j^*}\partial_iW \partial_{j^*}W^* + 2 \Re[
m( g^{ij^*} \partial_{j^*}K \partial_i W - 3W)
] + \mathcal{O}(m^2).
\end{align}
In our case even in the SUSY limit there is always a stabilizing higher-order piece to the potential coming from integrating out the adjoint, such that there is a finite minimum (this should be contrasted with, for instance, the pure ADS case where the minimum is at infinity). So we can expand around the SUSY minimum, for which $g^{i j^*}\partial_iW \partial_{j^*}W^*=0$. This implies the F-terms vanish $\partial_iW = 0$. It follows that the vacuum energy to leading order in $m$ is given by
\begin{align}
    V =& -6\Re[mW|_{m=0}] + \mathcal{O}(m^2), \label{eq:6mReW}
\end{align}
where $W|_{m=0}$ is the value of the superpotential evaluated at the $m=0$ SUSY minima. At no stage do we need to have knowledge of the K\"ahler potential. This makes our results much more robust. Solely with knowledge of the superpotential we can compute the leading-order vacuum energy.

The effective theories we use in Secs.~\ref{sec:ADS}--\ref{sec:FreeMag} are $\mathcal{N}=1$ descriptions with superpotentials $W_{\rm eff}$ that contain both explicit mass scales ({\it e.g.}\/, $\mu$ from the adjoint mass deformation) and holomorphic strong-coupling scales $\Lambda$ (and, in the magnetic regime, $\tilde\Lambda$). These parameters are precisely the quantities whose compensator dependence generates the leading $\theta$-dependent lifting of the supersymmetric vacuum degeneracy.

Concretely, in each regime we:
\begin{enumerate}
\item determine the supersymmetric vacua of the $\theta$-dependent effective theory ({\it i.e.}\/, the $m=0$ minima) from $W_{\rm eff}$, including the discrete branches associated with the holomorphic phases;
\item evaluate the AMSB contribution by inserting the corresponding $W_{\rm eff}$ into the universal expression~\eqref{eq:6mReW};
\item identify $V_{\rm AMSB}(\theta)$ among the competing supersymmetric branches, and determine $\theta_{\rm crit}$ by locating where the global minimum jumps.
\end{enumerate}

For completeness, AMSB also generates the standard loop-induced soft terms, expressed in terms of anomalous dimensions $\gamma_i$ and beta functions~\cite{pomarol_sparticle_1999,randall_out_1999,giudice_gaugino_1998}:
\begin{align}
A_{ijk}(\mu) &= - \frac{1}{2} (\gamma_{i} + \gamma_{j} + \gamma_{k})(\mu)\, m ,\\
m_{i}^{2}(\mu) &= - \frac{1}{4}\dot{\gamma}_{i}(\mu)\, m^{2} ,\\
m_{\lambda}(\mu) &= - \frac{\beta(g^{2})}{2g^{2}}(\mu)\, m .
\end{align}
In the present analysis, these loop-induced soft terms are not directly relevant. The mass term of the gauginos is of course not directly present in the scalar potential, and the two-loop scalar masses-squared are only relevant insofar as we restrict to $(N_f,N_c)$ such that the electric theory is asymptotically free and thus the scalar masses-squared are positive. Because we restrict to leading order in $m$ in theories that already have SUSY minima, these $\mathcal{O}(m^2)$ terms are negligible. At a supersymmetric stationary point, the linear variation associated with the $A$-terms is proportional to $\partial_iW$ and therefore vanishes in the $\mathcal O(m)$ vacuum energy. As such all of these can be neglected just by taking $m$ sufficiently small. This is particularly powerful in the case $N_f=N_c$, where there is no perturbative description and we cannot assume that anomalous dimensions are small.

\section{\texorpdfstring{$N_f<N_c$: ADS superpotential}{Nf<Nc}}\label{sec:ADS}

Since the deformation considered is irrelevant (even under the large
$\mu$ assumption), the low-energy theory in the supersymmetric limit
still develops the usual Affleck--Dine--Seiberg (ADS)
superpotential~\cite{ads_sqcd_1984,ads_chiral_1984,ads_four_dim_1985,
seiberg_exact_1994}. Integrating out the
adjoint produces the negative quartic deformation. Therefore, the
effective superpotential in terms of the quark and antiquark
superfields is
\begin{align}
W_{\rm eff}
={}&
(N_c-N_f)
\left(
\frac{\Lambda^{3N_c-N_f}}
{\det(\widetilde Q Q)}
\right)^{\frac{1}{N_c-N_f}}
-\frac{1}{\mu}
(\widetilde Q T^aQ)(\widetilde Q T^aQ).
\end{align}
Here
\begin{align}
\Lambda
=
|\Lambda|
\exp\left(\frac{i\theta}{3N_c-N_f}\right)
\end{align}
is the holomorphic scale of the $\mathcal N=1$ theory, and $\theta$
is in the same convention as Ref.~\cite{csaki_phase_2025}. Rewritten
in terms of the meson superfield, with the theta dependence explicit,
the superpotential is
\begin{align}
W_{\rm eff}
={}&
(N_c-N_f)
\left(
\frac{|\Lambda|^{3N_c-N_f}}{\det M}
\right)^{\frac{1}{N_c-N_f}}
\exp\left[
\frac{i(\theta+2\pi\ell)}{N_c-N_f}
\right]
\nonumber\\
&\quad
-\frac{1}{2\mu}\Tr(M^2)
+\frac{1}{2\mu N_c}(\Tr M)^2,
\qquad
\ell=0,\ldots,N_c-N_f-1.
\end{align}

It is useful to define
\begin{align}
d&\equiv N_c-N_f,
&
S_\ell
&\equiv
\left(
\frac{|\Lambda|^{3N_c-N_f}}{\det M}
\right)^{1/d}
\exp\left[
\frac{i(\theta+2\pi\ell)}{d}
\right].
\end{align}
In terms of $S_\ell$, the superpotential is
\begin{align}
W_{\rm eff}
=
dS_\ell
-\frac{1}{2\mu}
\left[
\Tr(M^2)-\frac{1}{N_c}(\Tr M)^2
\right].
\end{align}

The supersymmetric vacua are determined by
\begin{align}
\frac{\partial W_{\rm eff}}{\partial M}=0.
\end{align}
Using
\begin{align}
\frac{\partial S_\ell}{\partial M}
=
-\frac{1}{d}S_\ell M^{-1},
\end{align}
the matrix $F$-term equation is
\begin{align}
-S_\ell M^{-1}
-\frac{1}{\mu}
\left(
M-\frac{\Tr M}{N_c}\mathbf 1_{N_f}
\right)
=0.
\end{align}
Multiplying by $M$ gives
\begin{align}
M^2-\frac{\Tr M}{N_c}M
=
-\mu S_\ell\,\mathbf 1_{N_f}.
\label{eq:ADS-matrix-equation}
\end{align}

Since the right-hand side is proportional to the identity, $M$
obeys a quadratic equation and therefore has at most two distinct
eigenvalues. Writing these eigenvalues as $\alpha$ and $\beta$, with
multiplicities $r$ and $N_f-r$, respectively, gives
\begin{align}
M
&\sim
\operatorname{diag}
\left(
\alpha\,\mathbf 1_r,
\beta\,\mathbf 1_{N_f-r}
\right),
\\
\beta
&=
-\frac{N_c-r}{N_c-N_f+r}\,\alpha,
\qquad
r=0,\ldots,N_f.
\end{align}
The solutions labeled by $r$ and $N_f-r$ differ only by interchanging
the two eigenvalues, so one may restrict to
\begin{align}
0\leq r\leq\left\lfloor\frac{N_f}{2}\right\rfloor.
\end{align}

Defining
\begin{align}
L\equiv2N_c-N_f,
\qquad
d\equiv N_c-N_f,
\end{align}
the remaining $F$-term condition gives
\begin{align}
\alpha^L
=
(-1)^{N_c-r}\mu^d
\left(
\frac{d+r}{N_c-r}
\right)^{N_c-r}
|\Lambda|^{3N_c-N_f}e^{i\theta}.
\label{eq:ADS-a-root-equation}
\end{align}
For real positive $\mu$, the solutions across all ADS sheets may
therefore be written as
\begin{align}
\alpha_{r,j}
&=
A_r
\exp\left[
\frac{i}{L}
\left(
\theta+\pi(N_c-r)+2\pi j
\right)
\right],
\qquad
j=0,\ldots,L-1,
\\
A_r
&\equiv
\left[
\mu^d
\left(
\frac{d+r}{N_c-r}
\right)^{N_c-r}
|\Lambda|^{3N_c-N_f}
\right]^{1/L}.
\end{align}
Each solution is assigned to the corresponding original ADS sheet
$\ell$ by imposing the unpowered $F$-term equation.

Taking the trace of Eq.~\eqref{eq:ADS-matrix-equation} gives
\begin{align}
\Tr(M^2)-\frac{1}{N_c}(\Tr M)^2
=
-\mu N_fS_\ell.
\label{eq:ADS-trace-relation}
\end{align}
Consequently, the superpotential evaluated at the vacuum simplifies to
\begin{align}
W
&=
dS_\ell
-\frac{1}{2\mu}
\left[
\Tr(M^2)-\frac{1}{N_c}(\Tr M)^2
\right]
\nonumber\\
&=
\left(
d+\frac{N_f}{2}
\right)S_\ell
=
\frac{L}{2}S_\ell.
\end{align}
For an eigenvalue $\alpha$, Eq.~\eqref{eq:ADS-matrix-equation} gives
\begin{align}
S_\ell
=
-\frac{N_c-r}{\mu(d+r)}\alpha^2.
\end{align}
Thus, on the branch labeled by $(r,j)$,
\begin{align}
W_{r,j}
=
-\frac{L(N_c-r)}{2\mu(d+r)}\alpha_{r,j}^{\,2}.
\end{align}
Its real part is
\begin{align}
\Re W_{r,j}
=
-\frac{L(N_c-r)}{2\mu(d+r)}A_r^2
\cos\left[
\frac{
2\theta+2\pi(N_c-r)+4\pi j
}{L}
\right].
\label{eq:ADS-ReW}
\end{align}
At a supersymmetric vacuum, the leading AMSB contribution is therefore
\begin{align}
V_{\rm AMSB}^{(1)}
&=
-3mW_{r,j}+\mathrm{c.c.}
=
-6m\,\Re W_{r,j}
\nonumber\\
&=
\frac{3mL(N_c-r)}{\mu(d+r)}A_r^2
\cos\left[
\frac{
2\theta+2\pi(N_c-r)+4\pi j
}{L}
\right],
\end{align}
for real $m$.

The flavor-symmetric solution $M\propto\mathbf 1_{N_f}$ is the
$r=0$ solution (equivalently, the $r=N_f$ solution after interchanging
the eigenvalues). For $N_f\geq2$, the intermediate values of $r$
give additional mixed-eigenvalue supersymmetric branches.

\begin{figure}[t]
    \centering

    \begin{subfigure}[t]{0.48\linewidth}
        \centering
        \includegraphics[width=\linewidth]{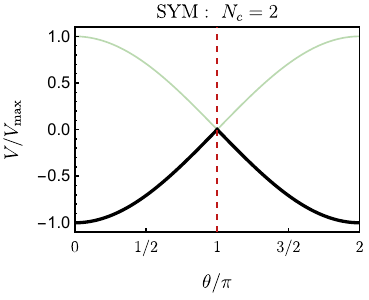}
        \caption{Pure $SU(2)$ SYM.}
        \label{fig:sym-nc2}
    \end{subfigure}
    \hfill
    \begin{subfigure}[t]{0.48\linewidth}
        \centering
        \includegraphics[width=\linewidth]{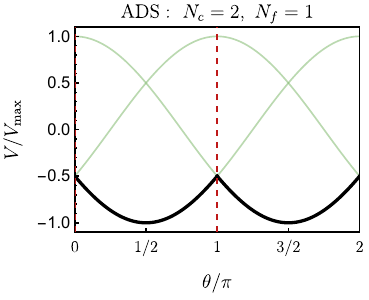}
        \caption{$SU(2)$ SQCD with $N_f=1$.}
        \label{fig:ads-nc2-nf1}
    \end{subfigure}

    \medskip

    \begin{subfigure}[t]{0.48\linewidth}
        \centering
        \includegraphics[width=\linewidth]{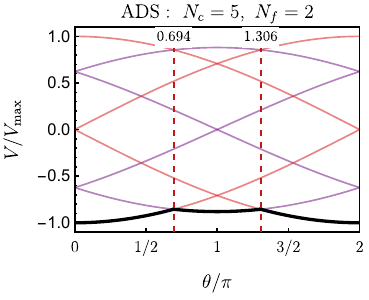}
        \caption{$SU(5)$ SQCD with $N_f=2$.}
        \label{fig:ads-nc5-nf2}
    \end{subfigure}

    \caption{Vacuum energies in pure SYM and in the ADS regime.
    Thin colored curves show the AMSB-lifted energies of the individual
    supersymmetric vacuum branches, while the thick black curve is their
    lower envelope and hence the physical vacuum energy. Red dashed lines
    mark the critical angles at which the global minimum changes branch.
    Pure SYM transitions only at $\theta=\pi$. In the ADS regime, the odd
    flavor count $N_f=1$ transitions at the endpoints
    $\theta_c/\pi=0,\,1$, while the even flavor count $N_f=2$ transitions
    at the generically noninteger values $\theta_c/\pi\simeq0.694,\,1.306$.
    Energies are normalized to $V_{\max}$ in each panel.}
    \label{fig:ads-plots}
\end{figure}

For $N_f>0$, the critical angles are
\begin{align}
\theta_{\rm crit}
=
\begin{cases}
0,\ \pi
\pmod{2\pi},
& N_f\ \text{odd},
\\[2mm]
\pm\delta_*
\pmod{2\pi},
& N_f\ \text{even},
\end{cases}
\label{eq:ADS-critical-angles}
\end{align}
where
\begin{align}
\delta_*
={}&
\left(
N_c-\frac{N_f}{2}
\right)
\arctan\left[
\frac{
\left(
\dfrac{N_c-\frac{N_f}{2}+1}
{N_c-\frac{N_f}{2}-1}
\right)^{\frac{1}{N_c-N_f/2}}
-\cos\left(
\dfrac{\pi}{N_c-N_f/2}
\right)
}{
\sin\left(
\dfrac{\pi}{N_c-N_f/2}
\right)
}
\right].
\end{align}
See Figure~\ref{fig:ads-plots}. For even $N_f$, the critical angles
are generically irrational multiples of $\pi$.

Formally setting $N_f=0$ in the ADS expressions produces the correct
functional dependence but overcounts the branches. The proper
description is pure $\mathcal N=1$ SYM, whose gaugino-condensation
superpotential is
\begin{align}
W_k
=
N_c|\Lambda|^3
\exp\left[
\frac{i(\theta+2\pi k)}{N_c}
\right],
\qquad
k=0,\ldots,N_c-1.
\end{align}
The leading AMSB vacuum energy is therefore
\begin{align}
V_k
=
-6mN_c|\Lambda|^3
\cos\left(
\frac{\theta+2\pi k}{N_c}
\right).
\end{align}
In the absence of flavors, the only imprint of integrating out the
adjoint is through the scale matching in
Eq.~\eqref{eq:scalematching}.

\section{\texorpdfstring{$N_f=N_c$: quantum-modified constraint}{Nf=Nc}}
\label{sec:QM}

For $N_f=N_c$, the moduli space of $\mathcal N=1$ SQCD obeys the
quantum-modified constraint
\begin{align}
\det M-\widetilde B B
=
\Lambda^{2N_c},
\end{align}
where $M^i{}_j$ is the meson superfield and $B,\widetilde B$ are the
baryon and antibaryon superfields, respectively
\cite{seiberg_exact_1994}; see
Ref.~\cite{intriligator_seiberg_lectures_1996} for a review. Writing
the theta dependence explicitly, we describe the constraint using a
Lagrange multiplier superfield $X$:
\begin{align}
W
={}&
X\left(
\det M-\widetilde B B
-|\Lambda|^{2N_c}e^{i\theta}
\right)
\nonumber\\
&\quad
-\frac{1}{2\mu}
\left[
\Tr(M^2)-\frac{1}{N_c}(\Tr M)^2
\right]
+A(\mu,\Lambda).
\label{eq:QMM-superpotential}
\end{align}
The field-independent holomorphic term $A(\mu,\Lambda)$ does not
affect the supersymmetric moduli space, but it can contribute to the
AMSB vacuum energy through the on-shell value of $W$.

Such a term must be included because the quantum-modified phase has
no perturbative description in terms of unconstrained IR composites.
An operator identity that vanishes when written in terms of the
elementary quarks need not vanish if the composite fields are first
treated as independent and the quantum-modified constraint is imposed
only afterward. Consequently, the constraint alone does not determine
the identity-operator part of the map from the UV deformation to the
IR composite theory. Dimensional analysis and holomorphy imply that
the leading possible ambiguity has the form
\begin{align}
A(\mu,\Lambda)
\sim
\frac{\Lambda^{2N_c}}{\mu^{2N_c-3}}.
\label{eq:QMM-contact-scaling}
\end{align}
For $N_c\geq3$, this begins at order $1/\mu^3$ and is subleading to
the order retained in our analysis. For $N_c=2$, however, it is of
order $\Lambda^4/\mu$ and must be included.

Since $A(\mu,\Lambda)$ is independent of the dynamical fields, it
does not enter the $F$-term equations. These are
\begin{align}
\det M-\widetilde B B
-|\Lambda|^{2N_c}e^{i\theta}
&=0,
&
X\widetilde B=XB
&=0,
\label{eq:QMM-constraint-baryon-F-terms}
\\
X\operatorname{adj}(M)
-\frac{1}{\mu}
\left(
M-\frac{\Tr M}{N_c}\mathbf 1_{N_c}
\right)
&=0.
\label{eq:QMM-M-F-term}
\end{align}
There are two classes of solutions.

\begin{enumerate}
\item
If $X=0$, Eq.~\eqref{eq:QMM-M-F-term} gives
\begin{align}
M
&=
\phi\,\mathbf 1_{N_c},
&
\phi^{N_c}-\widetilde B B
&=
|\Lambda|^{2N_c}e^{i\theta}.
\label{eq:QMM-X-zero-branch}
\end{align}
The quartic deformation vanishes identically on this branch, so its
on-shell superpotential is
\begin{align}
W_{X=0}=A(\mu,\Lambda).
\label{eq:QMM-W-X-zero-general}
\end{align}

\item
If $X\neq0$, the baryon $F$-term equations imply
\begin{align}
B=\widetilde B=0,
\qquad
\det M
=
|\Lambda|^{2N_c}e^{i\theta}.
\end{align}
Thus $M$ is invertible. Multiplying
Eq.~\eqref{eq:QMM-M-F-term} by $M$ gives
\begin{align}
M^2-\frac{\Tr M}{N_c}M
=
\mu X|\Lambda|^{2N_c}e^{i\theta}\mathbf 1_{N_c}.
\label{eq:QMM-quadratic-M}
\end{align}
The matrix $M$ therefore has at most two distinct eigenvalues, denoted
by $\alpha$ and $\beta$. If $\alpha$ has multiplicity $r$, with
$1\leq r\leq N_c-1$, the eigenvalues satisfy
\begin{align}
(N_c-r)\alpha+r\beta
&=0,
&
\alpha^r\beta^{N_c-r}
&=
|\Lambda|^{2N_c}e^{i\theta}.
\end{align}
Their solutions are
\begin{align}
\alpha_{r,k}
={}&
|\Lambda|^2
\left(
\frac{r}{N_c-r}
\right)^{\frac{N_c-r}{N_c}}
\exp\left[
\frac{i}{N_c}
\left(
\theta+\pi(N_c-r)+2\pi k
\right)
\right],
\nonumber\\
&\hspace{7cm}
k=0,\ldots,N_c-1,
\label{eq:QMM-alpha-solutions}
\\
\beta_{r,k}
={}&
-\frac{N_c-r}{r}\alpha_{r,k},
\\
X_{r,k}
={}&
\frac{N_c-r}{\mu r}
\frac{\alpha_{r,k}^2}
{|\Lambda|^{2N_c}e^{i\theta}}.
\end{align}
\end{enumerate}
The solutions labeled by $r$ and $N_c-r$ differ only by interchanging
the two eigenvalues. We now evaluate the on-shell superpotential,
treating separately the generic case $N_c\geq3$ and the special case
$N_c=2$.

\subsection{\texorpdfstring{Solutions for $N_c\geq3$}{Nc greater than or equal to 3}}

For $N_c\geq3$, Eq.~\eqref{eq:QMM-contact-scaling} shows that the
contact-term ambiguity is subleading at the order in $1/\mu$ retained
here, so we set $A=0$.

On the $X=0$ branch, Eq.~\eqref{eq:QMM-W-X-zero-general} then gives
\begin{align}
W_{X=0}=0.
\end{align}
This is the baryonic branch with vanishing leading AMSB energy.

On the $X\neq0$ branches, the constraint term vanishes on shell. Using
the eigenvalue relations in the quartic deformation gives
\begin{align}
W_{r,k}
={}&
-\frac{N_c|\Lambda|^4}{2\mu}
\left(
\frac{N_c-r}{r}
\right)^{\frac{2r-N_c}{N_c}}
\nonumber\\
&\times
\exp\left[
\frac{2i}{N_c}
\left(
\theta+\pi(N_c-r)+2\pi k
\right)
\right].
\label{eq:QMM-W-Nc-geq-3}
\end{align}
The corresponding leading AMSB energy is
\begin{align}
V_{r,k}(\theta)
={}&
-6m\,\Re W_{r,k}
\nonumber\\
={}&
\frac{3mN_c|\Lambda|^4}{\mu}
\left(
\frac{N_c-r}{r}
\right)^{\frac{2r-N_c}{N_c}}
\nonumber\\
&\times
\cos\left[
\frac{2}{N_c}
\left(
\theta+\pi(N_c-r)+2\pi k
\right)
\right].
\label{eq:QMM-V-Nc-geq-3}
\end{align}

For odd $N_c$, all values of $r$ lie on the same phase grid after a
relabeling of $k$, and the largest-amplitude branch controls the
physical vacuum. Its minimizing root changes at
\begin{align}
\theta_{\rm crit}
=
0,\ \pi
\pmod{2\pi}.
\end{align}
For even $N_c$, the two dominant branches lie on inequivalent phase
grids. Their crossings occur at
\begin{align}
\theta_{\rm crit}
=
\pm\frac{N_c}{2}
\arctan\left[
\frac{
\left(
\dfrac{N_c+2}{N_c-2}
\right)^{2/N_c}
-\cos\left(\dfrac{2\pi}{N_c}\right)
}{
\sin\left(\dfrac{2\pi}{N_c}\right)
}
\right]
\pmod{2\pi}.
\end{align}
Thus, for $N_c\geq3$,
\begin{align}
\theta_{\rm crit}
=
\begin{cases}
0,\ \pi
\pmod{2\pi},
& N_c\ \text{odd},
\\[3mm]
\displaystyle
\pm\frac{N_c}{2}
\arctan\left[
\frac{
\left(
\dfrac{N_c+2}{N_c-2}
\right)^{2/N_c}
-\cos\left(\dfrac{2\pi}{N_c}\right)
}{
\sin\left(\dfrac{2\pi}{N_c}\right)
}
\right]
\pmod{2\pi},
& N_c\ \text{even}.
\end{cases}
\label{eq:QMM-critical-angles}
\end{align}

\subsection{\texorpdfstring{Solutions for $N_c=2$}{Nc=2}}

For $N_c=2$, the contact term in
Eq.~\eqref{eq:QMM-contact-scaling} is of order $\Lambda^4/\mu$ and
cannot be neglected. Its origin can be seen explicitly by writing the
same UV deformation in the equivalent $SU(2)$ and $Sp(1)$ languages.

In the $SU(2)$ description, using the convention of
Ref.~\cite{csaki_phase_2025},
\begin{align}
W
=
\sqrt{2}\,\widetilde Q_i\Phi Q^i
+\mu\Tr(\Phi^2).
\end{align}
Integrating out the adjoint gives
\begin{align}
\Delta W_{SU(2)}
=
-\frac{1}{2\mu}
\left[
\Tr\bigl((Q^i\widetilde Q_i)^2\bigr)
-\frac12
\bigl(\Tr(Q^i\widetilde Q_i)\bigr)^2
\right].
\label{eq:QMM-SU2-UV-deformation}
\end{align}
Alternatively, introduce four $Sp(1)$ fundamentals by
\begin{align}
\widetilde Q_{i,\alpha}
&=
\epsilon_{\alpha\beta}\mathcal Q_{i+2}^{\beta},
&
Q^{i,\alpha}
&=
\mathcal Q_i^{\alpha},
\end{align}
and define
\begin{align}
h^{IJ}
=
\begin{pmatrix}
0_{2\times2}&\mathbf 1_{2\times2}\\
\mathbf 1_{2\times2}&0_{2\times2}
\end{pmatrix}.
\end{align}
The same adjoint coupling may then be written as
\begin{align}
\sqrt{2}\,\widetilde Q_i\Phi Q^i
=
\frac{\sqrt{2}}{2}
h^{IJ}\mathcal Q_I^\beta
(\epsilon\Phi)_{\beta\gamma}
\mathcal Q_J^\gamma,
\end{align}
and integrating out $\Phi$ gives
\begin{align}
\Delta W_{Sp(1)}
=
-\frac{1}{8\mu}
\Tr\left[
(h\mathcal Q\epsilon\mathcal Q)^2
\right].
\label{eq:QMM-Sp1-UV-deformation}
\end{align}

At the level of the elementary quarks,
Eqs.~\eqref{eq:QMM-SU2-UV-deformation} and
\eqref{eq:QMM-Sp1-UV-deformation} are identical. To express them in
terms of composites, define
\begin{align}
Q\widetilde Q
&=
M,
&
\mathcal Q\epsilon\mathcal Q
=
\widetilde M
&=
\begin{pmatrix}
b\epsilon&M\\
-M^T&\widetilde b\epsilon
\end{pmatrix}.
\end{align}
Using the Cayley--Hamilton identity
\begin{align}
\Tr(M^2)
=
(\Tr M)^2-2\det M,
\end{align}
one finds
\begin{align}
\Delta W_{Sp(1)}-\Delta W_{SU(2)}
=
\frac{1}{2\mu}
\left(
b\widetilde b-\det M
\right).
\label{eq:Sp-SU-difference}
\end{align}
The right-hand side vanishes identically when the composites are
written in terms of the elementary quarks. On the quantum-modified
moduli space, however,
\begin{align}
\det M-b\widetilde b
=
\Lambda^4,
\end{align}
so the two composite representatives differ by
\begin{align}
\Delta W_{Sp(1)}-\Delta W_{SU(2)}
=
-\frac{\Lambda^4}{2\mu}.
\end{align}
This is precisely a field-independent holomorphic term of the order
anticipated in Eq.~\eqref{eq:QMM-contact-scaling}.

Accordingly, the two composite descriptions must be supplemented by
contact terms:
\begin{align}
\Delta W_{SU(2)}
={}&
-\frac{1}{2\mu}\Tr(M^2)
+\frac{1}{4\mu}(\Tr M)^2
+A_{SU(2)}(\mu,\Lambda),
\\
\Delta W_{Sp(1)}
={}&
-\frac{1}{2\mu}\Tr(M^2)
+\frac{1}{4\mu}(\Tr M)^2
+\frac{1}{2\mu}
\left(
b\widetilde b-\det M
\right)
+A_{Sp(1)}(\mu,\Lambda).
\end{align}
They are equivalent on the quantum-modified moduli space provided
\begin{align}
A_{Sp(1)}
=
A_{SU(2)}
+\frac{\Lambda^4}{2\mu}.
\end{align}

Returning to the general $SU(2)$ composite description
\eqref{eq:QMM-superpotential}, the two classes of supersymmetric
solutions have
\begin{align}
W_{X=0}
&=
A(\mu,\Lambda),
\\
W_{X\neq0}
&=
A(\mu,\Lambda)
+\frac{|\Lambda|^4e^{i\theta}}{\mu}.
\label{eq:QMM-Nc2-branch-difference}
\end{align}
Thus the quantum-modified description determines the difference
\begin{align}
W_{X\neq0}-W_{X=0}
=
\frac{|\Lambda|^4e^{i\theta}}{\mu},
\end{align}
but not their common additive contribution.

\begin{figure}[t]
    \centering
    \begin{subfigure}[t]{0.48\linewidth}
        \centering
        \includegraphics[width=\linewidth]{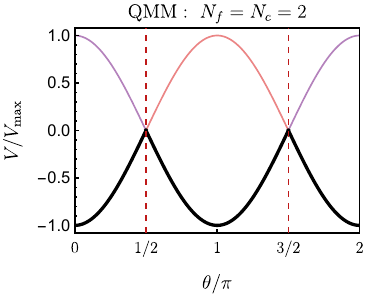}
        \caption{$N_f=N_c=2$.}
        \label{fig:qmm-nc2}
    \end{subfigure}
    \hfill
    \begin{subfigure}[t]{0.48\linewidth}
        \centering
        \includegraphics[width=\linewidth]{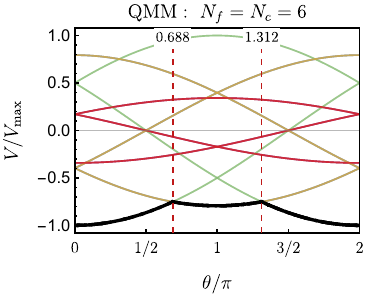}
        \caption{$N_f=N_c=6$.}
        \label{fig:qmm-nc6}
    \end{subfigure}

    \caption{Vacuum-branch structure in the quantum-modified moduli-space
    regime, $N_f=N_c$. Thin colored curves represent the energies of the
    competing supersymmetric branches after AMSB lifting, while the thick
    black curve is the physical vacuum energy obtained by minimizing over
    them. Red dashed lines indicate vacuum-branch transitions. For
    $N_c=2$, an unsuppressed contact term lifts the baryonic flat
    direction into a branch equal and opposite to the mesonic one, and the
    two cross at $\theta_c/\pi=1/2,\,3/2$; for $N_c=6$, the transitions
    occur at $\theta_c/\pi\simeq0.688,\,1.312$. Energies are normalized to
    $V_{\max}$ in each panel.}
    \label{fig:QMM-plots}
\end{figure}

We fix the latter by matching to the $\mathcal N=2$ result of
Ref.~\cite{csaki_phase_2025}. Agreement requires
\begin{align}
A(\mu,\Lambda)
=
-\frac{|\Lambda|^4e^{i\theta}}{2\mu}.
\end{align}
The two branches then have
\begin{align}
W_{X=0}
&=
-\frac{|\Lambda|^4e^{i\theta}}{2\mu},
&
W_{X\neq0}
&=
+\frac{|\Lambda|^4e^{i\theta}}{2\mu}.
\end{align}
Their leading AMSB energies are
\begin{align}
V_{X=0}(\theta)
&=
+\frac{3m|\Lambda|^4}{\mu}\cos\theta,
&
V_{X\neq0}(\theta)
&=
-\frac{3m|\Lambda|^4}{\mu}\cos\theta.
\end{align}
The physical vacuum energy is therefore
\begin{align}
V_{\min}(\theta)
=
-\frac{3m|\Lambda|^4}{\mu}
|\cos\theta|,
\end{align}
and the two branches cross at
\begin{align}
\theta_{\rm crit}
=
\frac{\pi}{2},\,
\frac{3\pi}{2}.
\end{align}
See Figure~\ref{fig:QMM-plots}.

\FloatBarrier
\section{\texorpdfstring{$N_f=N_c+1$: $s$-confinement}{Nf=Nc+1}}
\label{sec:sconf}

For $N_f=N_c+1$, the $\mathcal N=1$ SQCD theory is known to
$s$-confine~\cite{seiberg_exact_1994,
intriligator_seiberg_lectures_1996}. The AMSB deformation of
$s$-confining SQCD was studied in detail in
Ref.~\cite{de_lima_stolarski_2023}. The
low-energy superpotential is
\begin{align}
W_{\rm eff}
={}&
\frac{1}{\Lambda^{2N_c-1}}
\left(
\det M-\widetilde BMB
\right)
-\frac{1}{2\mu}
\left[
\Tr(M^2)-\frac{1}{N_c}(\Tr M)^2
\right].
\label{eq:sconf-superpotential}
\end{align}
Here and below, the theta dependence is contained in the holomorphic
scale,
\begin{align}
\Lambda^{2N_c-1}
=
|\Lambda|^{2N_c-1}e^{i\theta}.
\end{align}

The $F$-term equations are
\begin{align}
MB&=0,
&
\widetilde BM&=0,
\label{eq:sconf-baryon-F-terms}
\\
\operatorname{adj}(M)-B\widetilde B
-\frac{\Lambda^{2N_c-1}}{\mu}
\left(
M-\frac{\Tr M}{N_c}\mathbf 1_{N_c+1}
\right)
&=0.
\label{eq:sconf-M-F-term}
\end{align}
We find the following classes of solutions.

\begin{enumerate}
\item
Suppose that $\det M\neq0$. The baryon $F$-term equations then imply
\begin{align}
B=\widetilde B=0.
\end{align}
There are two types of full-rank solutions.

\begin{enumerate}
\item
The flavor-symmetric solution is
\begin{align}
M
=
\phi\,\mathbf 1_{N_f},
\end{align}
where Eq.~\eqref{eq:sconf-M-F-term} gives
\begin{align}
\phi^{N_c-1}
=
-\frac{\Lambda^{2N_c-1}}{\mu N_c}.
\label{eq:sconf-symmetric-root}
\end{align}

\item
The mixed-eigenvalue solutions take the form
\begin{align}
M
=
\begin{pmatrix}
\alpha \,\mathbf 1_r&0\\
0&\beta\,\mathbf 1_{N_f-r}
\end{pmatrix},
\qquad
2\leq r\leq N_c-1.
\end{align}
The two eigenvalues satisfy
\begin{align}
(N_c-r)\alpha+(r-1)\beta
&=
0,
\label{eq:sconf-eigenvalue-linear-relation}
\\
\alpha^{r-1}\beta^{N_f-r-1}
&=
-\frac{\Lambda^{2N_c-1}}{\mu}.
\label{eq:sconf-eigenvalue-product-relation}
\end{align}
Since $N_f=N_c+1$, the second equation may equivalently be written as
\begin{align}
\alpha^{r-1}\beta^{N_c-r}
=
-\frac{\Lambda^{2N_c-1}}{\mu}.
\end{align}
This class exists only for $N_c\geq3$. Solutions labeled by $r$ and
$N_f-r$ differ only by interchanging $\alpha$ and $\beta$, so inequivalent
solutions may be represented by
\begin{align}
2\leq r\leq
\left\lfloor\frac{N_c+1}{2}\right\rfloor.
\end{align}
\end{enumerate}

\item
Suppose that $\det M=0$ but $\Tr M\neq0$. The $F$-term equations imply
\begin{align}
M^2
=
\frac{\Tr M}{N_c}M.
\end{align}
Thus, $M$ has eigenvalues
\begin{align}
\phi
=
\frac{\Tr M}{N_c}
\end{align}
and zero, with respective multiplicities $N_c$ and one. We may
therefore write
\begin{align}
M
=
\phi(\mathbf 1_{N_f}-P),
\end{align}
where $P$ is an arbitrary rank-one projection. Such a projection can
be written as
\begin{align}
P=|v\rangle\langle w|,
\qquad
\langle w|v\rangle=1,
\end{align}
for a nonzero vector $|v\rangle\in\mathbb C^{N_f}$ and covector
$\langle w|\in(\mathbb C^{N_f})^*$. The baryon $F$-term equations
then require
\begin{align}
B
&=
b|v\rangle,
&
\widetilde B
&=
\widetilde b\langle w|,
\end{align}
and the remaining $M$ equation gives
\begin{align}
b\widetilde b
=
\phi^{N_c}
+\frac{\Lambda^{2N_c-1}}{\mu}\phi.
\label{eq:sconf-rank-Nc-baryon-condition}
\end{align}

\item
Suppose that $\det M=0$ and $\Tr M=0$. The $F$-term equations require
\begin{align}
\widetilde BB=0,
\qquad
M
=
-\frac{\mu}{\Lambda^{2N_c-1}}B\widetilde B.
\label{eq:sconf-nilpotent-branch}
\end{align}
It follows that
\begin{align}
M^2=0,
\qquad
\operatorname{rank}(M)\leq1.
\end{align}
This class includes branches with $M=B=0$ and $\widetilde B\neq0$,
as well as branches with $M=\widetilde B=0$ and $B\neq0$.
\end{enumerate}

Only the full-rank solutions have a nonzero on-shell superpotential.
Indeed, the baryon $F$-term equations imply $MB=0$ and hence
\begin{align}
\widetilde BMB=0
\end{align}
on every supersymmetric branch.

For the flavor-symmetric full-rank solution,
Eq.~\eqref{eq:sconf-symmetric-root} gives
\begin{align}
\frac{\det M}{\Lambda^{2N_c-1}}
=
\frac{\phi^{N_c+1}}{\Lambda^{2N_c-1}}
=
-\frac{\phi^2}{\mu N_c}.
\end{align}
Moreover,
\begin{align}
\Tr(M^2)-\frac{1}{N_c}(\Tr M)^2
=
-\frac{N_c+1}{N_c}\phi^2.
\end{align}
The on-shell superpotential is therefore
\begin{align}
W_{\rm sym}
=
\frac{N_c-1}{2\mu N_c}\phi^2.
\label{eq:sconf-W-symmetric}
\end{align}

For the mixed-eigenvalue solutions,
Eq.~\eqref{eq:sconf-eigenvalue-product-relation} implies
\begin{align}
\frac{\det M}{\Lambda^{2N_c-1}}
=
-\frac{\alpha\beta}{\mu}.
\end{align}
The eigenvalue relation also gives
\begin{align}
\Tr(M^2)-\frac{1}{N_c}(\Tr M)^2
=
-(N_c+1)\alpha\beta.
\end{align}
Consequently,
\begin{align}
W_{\rm mixed}
=
\frac{N_c-1}{2\mu}\alpha\beta.
\label{eq:sconf-W-mixed}
\end{align}

For the rank-$N_c$ solutions with $\det M=0$ and $\Tr M\neq0$,
\begin{align}
\Tr(M^2)
=
N_c\phi^2,
\qquad
\frac{1}{N_c}(\Tr M)^2
=
N_c\phi^2,
\end{align}
so every term in the on-shell superpotential vanishes:
\begin{align}
W=0.
\end{align}
For the nilpotent solutions with $\det M=\Tr M=0$, one has
\begin{align}
M^2=0,
\qquad
\widetilde BMB=0,
\end{align}
and again
\begin{align}
W=0.
\end{align}

We now write the nonzero branches with their theta dependence
explicit. For the flavor-symmetric branch,
Eq.~\eqref{eq:sconf-symmetric-root} gives
\begin{align}
\phi_k
=
\left(
\frac{|\Lambda|^{2N_c-1}}{\mu N_c}
\right)^{\frac{1}{N_c-1}}
\exp\left[
\frac{i}{N_c-1}
\left(
\theta+\pi+2\pi k
\right)
\right],
\qquad
k=0,\ldots,N_c-2.
\end{align}
Using Eq.~\eqref{eq:sconf-W-symmetric}, its real part is
\begin{align}
\Re W_{{\rm sym},k}
={}&
\frac{N_c-1}{2\mu N_c}
\left(
\frac{|\Lambda|^{2N_c-1}}{\mu N_c}
\right)^{\frac{2}{N_c-1}}
\nonumber\\
&\times
\cos\left[
\frac{2(\theta+\pi+2\pi k)}{N_c-1}
\right].
\label{eq:sconf-ReW-symmetric}
\end{align}

For a mixed-eigenvalue branch, Eq.~\eqref{eq:sconf-eigenvalue-linear-relation}
gives
\begin{align}
\beta
=
-\frac{N_c-r}{r-1}\alpha.
\end{align}
Substituting this into
Eq.~\eqref{eq:sconf-eigenvalue-product-relation} gives
\begin{align}
\alpha^{N_c-1}
=
(-1)^{N_c-r+1}
\left(
\frac{r-1}{N_c-r}
\right)^{N_c-r}
\frac{\Lambda^{2N_c-1}}{\mu}.
\end{align}
Thus
\begin{align}
\alpha_{r,k}
={}&
\left[
\left(
\frac{r-1}{N_c-r}
\right)^{N_c-r}
\frac{|\Lambda|^{2N_c-1}}{\mu}
\right]^{\frac{1}{N_c-1}}
\nonumber\\
&\times
\exp\left[
\frac{i}{N_c-1}
\left(
\theta+\pi(N_c-r+1)+2\pi k
\right)
\right],
\nonumber\\
&\hspace{7cm}
k=0,\ldots,N_c-2.
\end{align}
Using Eq.~\eqref{eq:sconf-W-mixed}, we obtain
\begin{align}
\Re W_{r,k}
={}&
-\frac{N_c-1}{2\mu}
\left(
\frac{|\Lambda|^{2N_c-1}}{\mu}
\right)^{\frac{2}{N_c-1}}
\left(
\frac{N_c-r}{r-1}
\right)^{\frac{2r-N_c-1}{N_c-1}}
\nonumber\\
&\times
\cos\left[
\frac{
2\bigl(\theta+\pi(N_c-r+1)+2\pi k\bigr)
}{N_c-1}
\right],
\nonumber\\
&\hspace{5cm}
2\leq r\leq
\left\lfloor\frac{N_c+1}{2}\right\rfloor.
\label{eq:sconf-ReW-mixed}
\end{align}

Combining all branches, the possible values of the real part of the
on-shell superpotential are
\begin{align}
\Re W
=
\begin{cases}
\displaystyle
\frac{N_c-1}{2\mu N_c}
\left(
\frac{|\Lambda|^{2N_c-1}}{\mu N_c}
\right)^{\frac{2}{N_c-1}}
\cos\left[
\frac{2(\theta+\pi+2\pi k)}{N_c-1}
\right],
&
k=0,\ldots,N_c-2,
\\[6mm]
\displaystyle
-\frac{N_c-1}{2\mu}
\left(
\frac{|\Lambda|^{2N_c-1}}{\mu}
\right)^{\frac{2}{N_c-1}}
\left(
\frac{N_c-r}{r-1}
\right)^{\frac{2r-N_c-1}{N_c-1}}
\cos\left[
\frac{
2\bigl(\theta+\pi(N_c-r+1)+2\pi k\bigr)
}{N_c-1}
\right],
&
\begin{gathered}
k=0,\ldots,N_c-2,\\[-1mm]
r=2,\ldots,
\left\lfloor\frac{N_c+1}{2}\right\rfloor,
\end{gathered}
\\[8mm]
0.&
\end{cases}
\label{eq:sconf-all-ReW}
\end{align}
The leading AMSB energy on every branch is
\begin{align}
V=-6m\,\Re W.
\end{align}

\begin{figure}[t]
    \centering
    \begin{subfigure}[t]{0.48\linewidth}
        \centering
        \includegraphics[width=\linewidth]{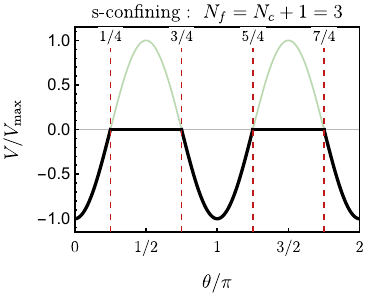}
        \caption{$N_c=2$, $N_f=3$.}
        \label{fig:sconf-nc2}
    \end{subfigure}
    \hfill
    \begin{subfigure}[t]{0.48\linewidth}
        \centering
        \includegraphics[width=\linewidth]{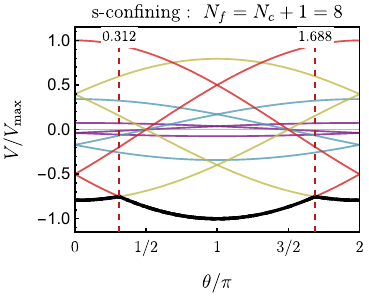}
        \caption{$N_c=7$, $N_f=8$.}
        \label{fig:sconf-nc7}
    \end{subfigure}

    \caption{Vacuum energies in the $s$-confining regime,
    $N_f=N_c+1$. Thin colored curves denote the AMSB-lifted candidate
    vacuum branches, while the thick black curve gives the global minimum.
    Red dashed lines mark the angles at which the minimizing branch changes.
    For $N_f=3$, transitions occur at
    $\theta_c/\pi=0.25,\,0.75,\,1.25,\,1.75$; for $N_f=8$, they occur at
    $\theta_c/\pi\simeq0.312,\,1.688$. Energies are normalized to
    $V_{\max}$ in each panel.}
    \label{fig:sconf-plots}
\end{figure}

The resulting critical angles are
\begin{align}
\theta_{\rm crit}
=
\begin{cases}
\displaystyle
\frac{\pi}{4}+\frac{\pi k}{2},
\qquad k=0,1,2,3,
& N_c=2,
\\[4mm]
\displaystyle
\frac{\pi}{2},\,\frac{3\pi}{2},
& N_c=3,
\\[4mm]
0,\,\pi,
& N_c\text{ even},\quad N_c\geq4,
\\[4mm]
\displaystyle
\pi\pm\delta_s
\pmod{2\pi},
& N_c\text{ odd},\quad N_c\geq5,
\end{cases}
\label{eq:sconf-critical-angles}
\end{align}
where
\begin{align}
\delta_s
={}&
\frac{N_c-1}{2}
\arctan\left[
\frac{
\left(
\dfrac{N_c+1}{N_c-3}
\right)^{\frac{2}{N_c-1}}
-\cos\left(
\dfrac{2\pi}{N_c-1}
\right)
}{
\sin\left(
\dfrac{2\pi}{N_c-1}
\right)
}
\right].
\end{align}
See Figure~\ref{fig:sconf-plots}. For odd $N_c\geq5$, the critical
angles are generically irrational multiples of $\pi$.

\FloatBarrier
\section{\texorpdfstring{$N_c+1<N_f<\frac32N_c$: Free Magnetic Phase}
{Nc+1<Nf<3Nc/2}}
\label{sec:FreeMag}

In this regime the IR description is the magnetic
$SU(N_f-N_c)$ theory, with magnetic quarks $q,\widetilde q$ and
gauge-singlet mesons $M$~\cite{seiberg_duality_1995}. The superpotential including the adjoint-induced
deformation is
\begin{align}
W
=
\frac{\widetilde qMq}{\mu_M}
-\frac{1}{2\mu}
\left[
\Tr(M^2)-\frac{1}{N_c}(\Tr M)^2
\right].
\end{align}
Defining
\begin{align}
X^i{}_j
=
\frac{1}{\mu_M}\widetilde q_jq^i,
\end{align}
the tree-level $F$-term equations are
\begin{align}
X-\frac{1}{\mu}
\left(
M-\frac{\Tr M}{N_c}\mathbf 1_{N_f}
\right)
&=0,
&
Mq=\widetilde qM&=0.
\end{align}
In gauge-invariant form, the quark equations imply $MX=XM=0$, and
hence
\begin{align}
M\left(
M-\frac{\Tr M}{N_c}\mathbf 1_{N_f}
\right)=0.
\end{align}

We find the following classes of rank-deficient solutions:
\begin{enumerate}
    \item If $\Tr M=0$, then
    \begin{align}
    M^2=0,
    \qquad
    X=\frac{M}{\mu}.
    \end{align}
    These branches have
    $\operatorname{rank}M\leq N_f-N_c$ and sufficiently many light
    magnetic flavors that no nonperturbative superpotential is
    generated. Since $\Tr(M^2)=\Tr M=\widetilde qMq=0$, they have
    \begin{align}
    W=0.
    \end{align}

    \item If $\Tr M\neq0$ and $M$ is rank deficient, then
    \begin{align}
    M=\phi P,
    \qquad
    P^2=P,
    \qquad
    \Tr P=N_c,
    \end{align}
    with
    \begin{align}
    X
    =
    -\frac{\phi}{\mu}
    \left(
    \mathbf 1_{N_f}-P
    \right).
    \end{align}
    The remaining $SU(N_f-N_c)$ theory has $N_f-N_c$ light flavors
    and obeys a quantum-modified constraint. This changes the baryonic
    moduli space but gives no on-shell contribution to the
    superpotential. Moreover,
    \begin{align}
    MX=0,
    \qquad
    \Tr(M^2)=\frac{1}{N_c}(\Tr M)^2,
    \end{align}
    so these branches also have
    \begin{align}
    W=0.
    \end{align}
\end{enumerate}
The remaining rank-deficient strata do not give additional finite
supersymmetric vacua. The rank-$(N_c-1)$ stratum is $s$-confining,
but its confining $F$-term equations are incompatible with the
complete $M$ equation. For
$N_c<\operatorname{rank}M<N_f$, the remaining magnetic theory
generates an ADS superpotential, and these fixed-rank strata run
toward the full-rank solutions.

If $M$ is invertible, then $q=\widetilde q=0$ and the low-energy
magnetic theory is pure $SU(N_f-N_c)$. Writing its gaugino condensate
as $S$, holomorphic scale matching gives
\begin{align}
S^{N_f-N_c}
=
(-1)^{N_f-N_c}
\frac{\det M}{\Lambda^{3N_c-N_f}},
\end{align}
where the magnetic matching scale $\mu_M$ has canceled. The
full-rank effective superpotential is
\begin{align}
W_{\rm eff}
=
-\frac{1}{2\mu}
\left[
\Tr(M^2)-\frac{1}{N_c}(\Tr M)^2
\right]
+(N_f-N_c)S.
\end{align}
Its $M$ equation is
\begin{align}
M^2-\frac{\Tr M}{N_c}M
-\mu S\,\mathbf 1_{N_f}
=0.
\label{eq:free-magnetic-quadratic}
\end{align}
Taking the trace of this equation gives
\begin{align}
\Tr(M^2)-\frac{1}{N_c}(\Tr M)^2
=
\mu N_fS,
\end{align}
and therefore every full-rank solution obeys
\begin{align}
W_{\rm on-shell}
=
-\frac{2N_c-N_f}{2}S.
\label{eq:free-magnetic-onshell}
\end{align}

Equation~\eqref{eq:free-magnetic-quadratic} admits a
flavor-symmetric full-rank solution,
\begin{align}
M=\phi\,\mathbf 1_{N_f}.
\end{align}
This solution must be treated separately from the mixed-eigenvalue
solutions below, whose parametrization assumes two distinct
eigenvalues. For the flavor-symmetric solution,
\begin{align}
S
=
-\frac{N_f-N_c}{\mu N_c}\phi^2,
\end{align}
and scale matching gives
\begin{align}
\phi^{2N_c-N_f}
=
\left(
\frac{N_f-N_c}{N_c}
\right)^{N_f-N_c}
\frac{\Lambda^{3N_c-N_f}}{\mu^{N_f-N_c}}.
\end{align}
Taking
\begin{align}
\Lambda^{3N_c-N_f}
=
|\Lambda|^{3N_c-N_f}e^{i\theta},
\end{align}
the roots are
\begin{align}
\phi_j
={}&
\left[
\left(
\frac{N_f-N_c}{N_c}
\right)^{N_f-N_c}
\frac{|\Lambda|^{3N_c-N_f}}{\mu^{N_f-N_c}}
\right]^{\frac{1}{2N_c-N_f}}
\nonumber\\
&\times
\exp\left[
\frac{i(\theta+2\pi j)}{2N_c-N_f}
\right],
\qquad
j=0,\ldots,2N_c-N_f-1.
\end{align}
Using Eq.~\eqref{eq:free-magnetic-onshell}, the corresponding real
parts of the superpotential are
\begin{align}
\Re W_{{\rm sym},j}
={}&
\frac{(2N_c-N_f)(N_f-N_c)}{2\mu N_c}
\left[
\left(
\frac{N_f-N_c}{N_c}
\right)^{N_f-N_c}
\frac{|\Lambda|^{3N_c-N_f}}{\mu^{N_f-N_c}}
\right]^{\frac{2}{2N_c-N_f}}
\nonumber\\
&\times
\cos\left[
\frac{2(\theta+2\pi j)}{2N_c-N_f}
\right].
\label{eq:free-magnetic-symmetric-ReW}
\end{align}

Apart from this flavor-symmetric branch, the remaining full-rank
solutions have two distinct nonzero eigenvalues $\alpha$ and $\beta$, with
multiplicities $N_f-N_c+r$ and $N_c-r$, respectively, where
\begin{align}
1\leq r\leq
\left\lfloor\frac{2N_c-N_f}{2}\right\rfloor.
\end{align}
The eigenvalue equations give
\begin{align}
\alpha
=
-\frac{r}{2N_c-N_f-r}\,y,
\qquad
S
=
\frac{r}{\mu(2N_c-N_f-r)}\,\beta^2.
\end{align}
Combining these equations with scale matching gives
\begin{align}
\beta^{2N_c-N_f}
=
(-1)^r
\left(
\frac{2N_c-N_f-r}{r}
\right)^r
\frac{\Lambda^{3N_c-N_f}}{\mu^{N_f-N_c}}.
\end{align}
The corresponding roots are
\begin{align}
\beta_{r,j}
={}&
\left[
\left(
\frac{2N_c-N_f-r}{r}
\right)^r
\frac{|\Lambda|^{3N_c-N_f}}{\mu^{N_f-N_c}}
\right]^{\frac{1}{2N_c-N_f}}
\nonumber\\
&\times
\exp\left[
\frac{i\left(\theta+\pi r+2\pi j\right)}
{2N_c-N_f}
\right],
\qquad
j=0,\ldots,2N_c-N_f-1.
\end{align}
Using Eq.~\eqref{eq:free-magnetic-onshell}, the real parts of the
on-shell superpotential are
\begin{align}
\Re W_{r,j}
={}&
-\frac{2N_c-N_f}{2}
\frac{
|\Lambda|^{\frac{2(3N_c-N_f)}{2N_c-N_f}}
}{
\mu^{\frac{N_f}{2N_c-N_f}}
}
\left(
\frac{r}{2N_c-N_f-r}
\right)^{\frac{2N_c-N_f-2r}{2N_c-N_f}}
\nonumber\\
&\times
\cos\left[
\frac{
2\theta+2\pi r+4\pi j
}{
2N_c-N_f
}
\right].
\label{eq:free-magnetic-mixed-ReW}
\end{align}
The leading AMSB energy on every branch is
\begin{align}
V=-6m\,\Re W.
\end{align}

The flavor-symmetric branch never controls the physical vacuum
energy. If $2N_c-N_f$ is even, it has the same phase grid as the
$r=(2N_c-N_f)/2$ mixed branch, but its amplitude relative to that
branch is
\begin{align}
\left(
\frac{N_f-N_c}{N_c}
\right)^{\frac{N_f}{2N_c-N_f}}
<\frac18.
\end{align}
It is therefore pointwise subdominant. If $2N_c-N_f$ is odd, the
dominant mixed branch has
\begin{align}
r=\frac{2N_c-N_f-1}{2}.
\end{align}
The maximum over its discrete roots is everywhere larger than
$2^{-4/3}$ times the common overall scale in
Eq.~\eqref{eq:free-magnetic-mixed-ReW}, whereas the symmetric branch
is smaller than one eighth of the same scale. Thus the symmetric
branch is again subdominant and does not modify the critical angles.

If $2N_c-N_f$ is odd, all mixed-eigenvalue branches have the same
phase grid after relabeling their roots, and the largest branch has
\begin{align}
r=\frac{2N_c-N_f-1}{2}.
\end{align}
The maximizing root changes at
\begin{align}
\theta_{\rm crit}
=
0,\,
\pi
\pmod{2\pi}.
\end{align}
These transitions exchange discrete roots of the same full-rank
branch.

\begin{figure}
    \centering
    \includegraphics[width=0.6\linewidth]{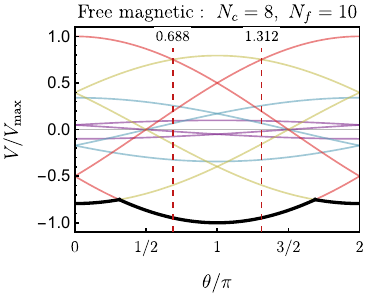}
    \caption{Vacuum energies in the free magnetic phase for $N_c=8$ and
    $N_f=10$. Thin colored curves denote the competing full-rank branches
    — both the mixed-eigenvalue branches and the flavor-symmetric branch
    $M=\phi\mathbf{1}_{N_f}$, which is always subdominant but is a genuine
    supersymmetric branch — while the thick black curve gives the global
    minimum. Red dashed lines mark the branch-changing transitions at
    $\theta_c/\pi\simeq0.688,\,1.312$. Energies are normalized to
    $V_{\max}$.}
    \label{fig:freeMag_plots}
\end{figure}

If $2N_c-N_f$ is even, there are two inequivalent phase grids. The
only branches that can control the physical vacuum energy have
\begin{align}
r=\frac{2N_c-N_f}{2},
\qquad
r=\frac{2N_c-N_f}{2}-1.
\end{align}
Their magnitude ratio is
\begin{align}
\frac{
\left|W_{\frac{2N_c-N_f}{2}-1}\right|
}{
\left|W_{\frac{2N_c-N_f}{2}}\right|
}
=
\left(
\frac{2N_c-N_f-2}{2N_c-N_f+2}
\right)^{\frac{2}{2N_c-N_f}}.
\end{align}
The critical angles are
\begin{align}
\theta_{\rm crit}^{\pm}
={}&
\pm\frac{2N_c-N_f}{2}
\arctan\left[
\frac{
\left(
\frac{2N_c-N_f+2}{2N_c-N_f-2}
\right)^{\frac{2}{2N_c-N_f}}
-\cos\left(\frac{2\pi}{2N_c-N_f}\right)
}{
\sin\left(\frac{2\pi}{2N_c-N_f}\right)
}
\right]
\pmod{2\pi}.
\end{align}
These transitions exchange the two dominant mixed-eigenvalue
full-rank branches. As in the other regimes,
$\theta_{\rm crit}/\pi$ is generically irrational. See Figure~\ref{fig:freeMag_plots}.

\FloatBarrier
\section{Conclusion}\label{sec:conclusion}

In this paper we studied $\theta$-dependence and the associated phase structure in an infinite family of $\mathcal N=1$ cousins of QCD. The starting point is $\mathcal N=1$ SQCD$(N_c,N_f)$ deformed by a quartic meson operator, $\Delta W=-\frac{1}{\mu}(\tilde{Q}T^a Q)(\tilde{Q} T^a Q)=-\frac{1}{2\mu}\Tr(M^2)+\frac{1}{2\mu N_c}(\Tr M)^2$, and by anomaly-mediated supersymmetry breaking (AMSB). These theories arise as $\mathcal N=1$ effective descriptions of an $\mathcal N=2$ parent (an $SU(N_c)$ theory with $N_f$ hypermultiplets) after giving a mass to the adjoint and integrating it out. The key simplification of our approach is that we never need to solve the full $\mathcal N=2$ special-geometry problem: we work directly with regime-by-regime $\mathcal N=1$ effective superpotentials and determine the $\theta$-dependence of the vacuum energy through the universal AMSB spurion prescription.

Our central mechanism is that, at $m=0$, the effective superpotentials exhibit a discrete set of supersymmetric branches distinguished by the holomorphic phases of the strong-coupling scales (equivalently, by discrete choices of the $N_c$- and $N_f$-dependent roots). AMSB lifts the degeneracy between these branches. Because supersymmetric vacua have vanishing $F$-term potential, the leading $\theta$-dependence comes entirely from the compensator-induced $m$-dependent terms. As a result, in each regime the vacuum energy takes a branch-selecting cosine form, and the ``preferred'' vacuum changes at special values $\theta_{\rm crit}$, signaling a vacuum-branch phase transition.

The critical-angle pattern is controlled by discrete vacuum-branch data encoded in the holomorphic phases of the strong-coupling scales, with AMSB selecting among the branches through the compensator-induced energy splitting. This structure accounts for phase transitions at fractional and even irrational values of $\theta/\pi$, rather than only at the conventional CP-symmetric points. It also explains our restriction to $N_f/N_c<3/2$: available treatments of the conformal window rely on a large-$N_c$ limit at fixed $N_f/N_c$, which does not retain the definite parities of $N_c$, $N_f$, and $2N_c-N_f$ that determine our critical angles. Our method reproduces the vacuum energy as a function of $\theta$ obtained for $N_c=2$ using the full $\mathcal N=2$ analysis of Ref.~\cite{csaki_phase_2025}. This agreement is consistent with the holomorphic continuity between the strongly coupled dyon regime considered there and the regime in which our effective $\mathcal N=1$ description is manifestly valid. Our approach is independent of the K\"ahler potential and as such is robust against any unknown features thereof. 

Motivated by holomorphic continuity and the explicit $N_c=2$ comparison, we formulate the following conjecture. The two computations correspond to different orderings of the same operations: (1) flow to low energy within $\mathcal N=2$, then break to $\mathcal N=1$ and add AMSB; or (2) break to $\mathcal N=1$ at high energy and add AMSB, then flow to low energy. We conjecture that these orderings yield the same IR theory for all $N_f$ and $N_c$, provided AMSB remains a sufficiently small perturbation. If this conjecture holds, our simpler $\mathcal N=1$ method provides a practical tool for determining the features of the vacua in the infinite family of theories for which the full $\mathcal N=2$ analysis is intractable.

Several directions merit further study. It would be valuable to (i) explicitly verify the vacuum energy results against the full $\mathcal N=2$ computation for additional $(N_c,N_f)$, most naturally at $N_c=3$ where a direct comparison is in principle possible (building on~\cite{dhoker_multimonopole_2021, dhoker_strong_coupling_2022}); (ii) determine the order and universality class of the resulting transitions by analyzing fluctuations near $\theta_{\rm crit}$; and (iii) understand whether and how these critical angles connect to the non-supersymmetric limit envisioned in the broader AMSB program.

\section*{Acknowledgments}
\hm{We thank Csaba Cs\'aki and Ofri Telem for encouragements. }
The work of HM was supported by the NSF grant PHY-2515115, by the U.S. Department of Energy (DE-AC02-05CH11231), by the JSPS Grant-in-Aid for Scientific Research JP23K03382, MEXT Grant-in-Aid for Transformative Research Areas (A) 26H00401, 26A204, 26H00403, Hamamatsu Photonics, K.K, Tokyo Dome Corporation, and by the World Premier International Research Center Initiative (WPI) MEXT, Japan. The work of BN was supported by the NSF grant PHY-2515115 and the UC Dissertation-Year Fellowship.

\bibliographystyle{JHEP}
\bibliography{refs.bib}

\end{document}